\documentclass[fleqn,usenatbib]{mnras}

\usepackage{newtxtext,newtxmath}

\usepackage[T1]{fontenc}

\DeclareRobustCommand{\VAN}[3]{#2}
\let\VANthebibliography\thebibliography
\def\thebibliography{\DeclareRobustCommand{\VAN}[3]{##3}\VANthebibliography}

\usepackage{graphicx}	%
\usepackage{amsmath}	%
\usepackage{siunitx}  %
\usepackage{subfig}

\newcommand{\msol}{\text{M}_\odot}
\newcommand{\kms}{km.s^{-1}}
\newcommand{\mass}{\text{M}}

\newcommand{\rms}[1]{\ensuremath{_{\text{#1}}}}

\newcommand{\atom}[2]{$^{#2}\text{#1}$}

\newcommand{\al}{\atom{Al}{26}}
\newcommand{\fe}{\atom{Fe}{60}}

\newcommand{\amuse}{\texttt{AMUSE}}
\newcommand{\bhtree}{\texttt{BHTree}}
\newcommand{\seba}{\texttt{SeBa}}

\title[SLR enrichment from AGB interlopers]{Short-Lived Radioisotopic enrichment from AGB interlopers in low-mass star-forming regions}

\author[J. W. Eatson et al.]{
Joseph W. Eatson$^{1,2}$\thanks{E-mail: \texttt{\href{mailto:j.w.eatson@sheffield.ac.uk}{j.w.eatson@sheffield.ac.uk}}} and 
Richard J. Parker$^{1}$\thanks{Royal Society Dorothy Hodgkin Fellow}
\\
$^{1}$Astrophysics Research Cluster, School of Mathematical and Physical Sciences, The University of Sheﬃeld, Hounsfield Road, Sheﬃeld, S3 7RH, UK \\
$^{2}$Lunar \& Planetary Laboratory, University of Arizona, AZ 85721, USA
}

\pubyear{2023}

\begin{document}
\label{firstpage}
\pagerange{\pageref{firstpage}--\pageref{lastpage}}
\maketitle

\begin{abstract}
The decay of Short-Lived Radioisotopes (SLRs) can be a significant source of heating early in protoplanetary systems, though how a protoplanetary disk becomes enriched with these SLRs far above the galactic background level remains an open question.
Observational evidence suggests that this enrichment occurs during the period from when the disk forms to when it progresses into a protoplanetary system, and is homogenous throughout the resultant planetary system.
Whilst SLRs such as \al{} and \fe{} can be injected into disks through interaction with Wolf-Rayet winds and supernovae, these outflows can disrupt disks, and in the case of low-mass star-forming regions high-mass stars may not form at all.
Recent research has determined the existence of AGB ``interlopers'', Asymptotic Giant Branch stars that pass close to or through star-forming regions that could be an alternative source of SLR enrichment to WR winds and supernovae.
In this paper we study the effect of AGB interlopers on star-forming regions from a dynamical perspective, determining the enrichment amount of \al{} and \fe{} in disks within small clusters via numerous $N$-body simulations via a parameter space exploration.
We find that enrichment is widespread from AGB stars, with efficient enrichment dependent on the time at which the interloper intersects the star-forming region.
Velocity is a factor, though interlopers travelling at \SI{30}{km.s^{-1}} are capable of enriching many disks in a star-forming region assuming they encounter a disk when the interloper is more evolved.

\end{abstract}

\begin{keywords}
exoplanets -- planets and satellites: terrestrial planets -- planets and satellites: interiors --  planets and satellites: composition -- planets and satellites: formation
\end{keywords}

\section{Introduction}

Short-lived radioisotopes (SLRs) such as \al{} and \fe{} are isotopes with half-lives on the order of \SI{1}{Myr}.
These isotopes have had a marked impact on the evolution of our Solar System, with their stable decay products being found throughout the Solar System within chondritic meteorites \citep{thraneExtremelyBriefFormation2006,tangAbundanceDistributionOrigin2012,mishraAbundance60FeInferred2016,davisShortLivedNuclidesEarly2022}.
The decay heating from these isotopes provided the bulk of heating in the early Solar System, speeding up the process of elemental stratification in planetesimals and causing the outgassing of volatiles such as H\textsubscript{2}O \citep{hirschmannEarlyVolatileDepletion2021}.
Outside the Solar System, the presence of these isotopes could cause desiccation of planetesimals, leading to a larger number of water-poor and rocky worlds entirely covered in an ocean.
It has been found that in the case of planetesimals with similar bulk compositions to the Solar System that significant desiccation from \al{} heating begins to occur at $\sim 0.1\times$ Solar \al{} enrichment \citep{eatsonDevolatilizationExtrasolarPlanetesimals2024,lichtenbergWaterBudgetDichotomy2019}.
Future exoplanet-oriented space telescopes such as the PLATO mission could potentially determine population statistics for desiccated worlds and ocean worlds via radius and bulk density estimations \citep{lichtenbergWaterBudgetDichotomy2019}.

Due to the short-lived nature of these isotopes and homogeneous distribution throughout the Solar System, it is likely that SLRs are injected into a star systems protoplanetary disk within a few Myr of formation from a source within the star-forming region \citep{trappitschSolarCosmicrayInteraction2015,deschShortLivedRadionuclidesMeteorites2022}.
The primary mechanisms of SLR injection that have been studied are through early-type stars, particularly ones with a Wolf-Rayet phase.
Evolved early-type stars can dredge-up substantial quantities of \al{} from their cores and eject them into the surrounding medium through their dense, fast winds \citep{arnould1997short}.
Once the massive star undergoes supernova, substantial quantities of both \al{} \& \fe{} are ejected into the surrounding medium, providing a significant enrichment mechanism for \fe{} \citep{limongiPresupernovaEvolutionExplosive2018,limongiNucleosynthesis26Al60Fe2006}.
Previous simulations have shown that SNe and Wolf-Rayet winds can result in the observed Solar System levels of enrichment in a significant fraction of disks, however, the Solar System is on the upper end of both \al{} and \fe{} enrichment \citep{eatsonUnifiedInjectionModel2024,parkerShortlivedRadioisotopeEnrichment2023}.

Wind and SNe injection are relatively straightforward explanations of SLR enrichment; however, there are some caveats.
Most importantly, wind-based injection is heavily biased in favour of producing lighter elements such as \al{}, leading to \fe{} enrichment being dominated by SNe-based injection \citep{limongiNucleosynthesis26Al60Fe2006}.
As noted in our previous papers and stellar evolution simulations of SNe SLR yields \citep{eatsonUnifiedInjectionModel2024,parkerShortlivedRadioisotopeEnrichment2023,limongiPresupernovaEvolutionExplosive2018}, stars with a mass of $\mass_\star < 25\,\si{\msol}$ undergo supernovae, with more massive stars directly collapsing into black holes.
Assuming that stars form in a star-forming region at roughly the same time, this puts the first supernovae occurring in the region at $\approx \SI{8}{Myr}$ after initial collapse, at which point most disks have progressed to protoplanetary systems, and would not efficiently sweep up SLRs.
Additionally, interactions with winds and photoionising radiation from nearby massive stars would cause significant disruption of gas within a disk, which would inhibit the formation of gas giants \citep{nicholsonRapidDestructionProtoplanetary2019,parkerFarExtremeUltraviolet2021,concha-ramirezExternalPhotoevaporationCircumstellar2019}.
$N$-body simulations incorporating a photodissociation model show that even limited close passes can significantly disrupt a disks typical formation
\citep{patelPhotoevaporationEnrichmentCradle2023a}.
However, cloud shielding could protect the nascent protoplanetary disk from external photoevaporation during the crucial early formation stage \citep{qiaoPlanetFormationPebble2023,colemanHowMicrophysicalProperties2025}.
Stellar feedback and shocks from the SNe would also contribute to significant disk disruption as well.
Finally, due to the rarity of high-mass stars, smaller, lower mass star-forming regions may lack stars of sufficient mass entirely \citep{nicholsonSupernovaEnrichmentPlanetary2017}.

Cosmic ray spallation from the planetary systems star could also replenish SLRs, though this replenishment only covers \al{} and is not sufficient and would be inhomogeneous within the disk, which is not observed in our Solar System \citep{trappitschSolarCosmicrayInteraction2015}.

SLRs can also be inherited from the disk's parent molecular cloud, but would likely result in insufficient SLR quantities due to their rapid decay timescale relative to the collapse of the molecular cloud.
Galactic chemical evolution, however, can also be used to explain the presence of low-concentration SLRs such as \atom{Mn}{53}, \atom{Pd}{107} and \atom{Hf}{182}.
Sequential star formation-based enrichment \citep{gounelleOriginShortlivedRadionuclides2008,gounelleSolarSystemGenealogy2012a,gounelleAbundance26Rich2015} could result in higher levels of enrichment over the Galactic background. However, recent stellar evolution models \citep{limongiPresupernovaEvolutionExplosive2018,2020ApJ...890...51E} suggest that the timescales involved would be too long and the requirement for subsequent generations to form one or more Wolf-Rayet stars would mean that such enrichment would be improbable, even for SLRs with a longer half-life such as \fe{} \citep{parkerDidSolarSystem2016}.

Instead, an alternative means of production of SLRs must be considered: interloping Asymptotic Giant Branch stars.
In an AGB star \al{} is readily produced through cool bottom processes \citep{nollettCoolBottomProcesses2003}, and \fe{} is transferred to the star's envelope through dredge-up. These fusion products can be ejected from the star through the stellar wind into the surrounding medium, causing SLR enrichment \citep{trigo-rodriguezRoleMassiveAGB2009,lugaro26Al60FeYields2008}.
\al{} yields from an AGB star are equivalent to a WR star, while having a markedly higher \fe{} yield \citep{Karakas14,karakasStellarYieldsMetalRich2016,lugaroRadioactiveNucleiCosmochronology2018}.
Furthermore, these less disruptive winds and lower photo-ionising flux from the AGB star will also impact disk evolution much less than WR stars and supernovae.
The possibility of pollution from an AGB star as an enrichment mechanism for the Solar System has been discussed previously in the literature 
\citep{bussoNucleosynthesisAsymptoticGiant1999}. Although it avoids the issues with theories for SLR enrichment involving massive stars, there is still the caveat that the probability of such an encounter would be low, especially for an intermediate-mass AGB \citep{kastnerObservationalEstimateProbability1994}.

Despite this, we are re-considering this mechanism as recent strides in astrometry with \emph{Gaia} have been able to distinguish interloping stars in star-forming regions from bona fide members \citep{schoettlerDoubleTroubleGaia2021}.
In particular, \citet{parkerIsotopicEnrichmentPlanetary2023} report the discovery of an AGB interloper that passed through the young star-forming region NGC 2264 as observed by Gaia, and demonstrated that this interaction could enrich a protoplanetary disk to SLR levels observed in the early Solar System.
Interloper candidates have been identified in other star-forming regions (Dr. C. Schoettler, private communication), and these serendipitous discoveries could mean that interlopers are more common than once thought.  Despite this, a calculation for the overall probability of an AGB interloper event has yet to be performed.

Other SLRs, such as \atom{Mn}{53}, \atom{Pd}{107} and 
\atom{Hf}{182} were also present in the early Solar System, though in much lower quantities relative to the production of \al{} or \fe{}.
For the Solar System, these other SLRs can be explained by galactic chemical evolution \citep{truemanGalacticChemicalEvolution2022}, and AGB interloper enrichment is not capable of self-consistently explaining all SLR enrichment. SLR enrichment of planetary systems is likely to be common \citep{lichtenbergIsotopicEnrichmentForming2016,curryPrevalenceShortlivedRadioactive2022}, but the destructive effects of photoionising radiation from massive stars may preclude gas giant formation \citep{patelPhotoevaporationEnrichmentCradle2023a}, and therefore we require an alternative enrichment scenario for the Solar System.
Another important consideration for the probability of an interloping event occurring is the relatively short length of the AGB phase relative to the lifespan of a low-mass main sequence star; typically this phase is of the order of a few Myr, meaning an AGB candidate could pass entirely through a star-forming region without entering the phase, leading to no enrichment.
Further considerations would have to be made in regard to the intercept velocity of the interloper, typical ``runaway'' ($>30\,\si{\kms}$) speeds may significantly curtail enrichment, while ``walkaway'' speeds ($5 - 30\,\si{\kms}$) may be more conducive to enrichment.

In this paper we present results from a series of simulations in order to determine the range of viable intercept velocities and AGB phase time sensitivity for significant SLR enrichment due to interloping stars, with a focus on intercept velocities at or below the galactic velocity dispersion -- so-called ``walkaway'' stars.
We expand on previous literature by examining the dynamics between the interloper and star-forming region. We also expand on the model detailed in our previous paper on SLR enrichment \citep{eatsonUnifiedInjectionModel2024} by accounting for SLRs lingering in the simulation, rather than having an ``enrichment zone'' around the SLR sources.
While there are a variety of studies that focus on the feasibility of AGB stellar nucleosynthesis for SLR enrichment \citep{bussoNucleosynthesisAsymptoticGiant1999,wasserburgIntermediatemassAsymptoticGiant2017}, and there are studies that focus on the probability of interactions between a star-forming region and an interloping AGB star \citep{kastnerObservationalEstimateProbability1994}; we instead focus on the feasibility from a dynamical perspective within the star-forming region itself.
As there are observed examples of an interloping AGB star \citep{parkerIsotopicEnrichmentPlanetary2023} it is of interest to determine the likelihood that protoplanetary disks within a star-forming region would undergo significant enrichment.
It is also of interest to consider how this likelihood changes based on interloper velocity and intercept distance, as well as the size and density of the star-forming region.
The paper is organised as follows. In Section \ref{sec:methodology} we discuss the methodology of this paper, in particular, how we expand on the model used in \citet{eatsonUnifiedInjectionModel2024} and covering how interloper enrichment is performed in our $N$-body simulations, and how we manage interlopers with a high intercept velocity with the star-forming region.
Our results for our simulation sets are detailed in Section \ref{sec:results}, and we provide  a brief discussion in Section \ref{sec:discussion}. We conclude in Section~\ref{sec:conclusion}.

\section{Methodology}
\label{sec:methodology}

The simulation code is written in Python, and uses the \amuse{} framework for its stellar evolution and $N$-body routines \citep{zwartAstrophysicalRecipesArt2018,portegieszwartMultiphysicsSimulationsUsing2013,pelupessyAstrophysicalMultipurposeSoftware2013}.
The problem is split into a series of sub-steps in order to synchronize the various codes.
For $N$-body simulation the \bhtree{} code \citep{barnesHierarchicalLogForcecalculation1986} was used, for stellar evolution \seba{} \citep{portegieszwartPopulationSynthesisHighmass1996,toonenSupernovaTypeIa2012} was utilised.

\subsection{Star-forming regions}

Observations \citep{Andre14,Hacar17} and simulations \citep{Schmeja06,Bate12,Girichidis12} of the earliest stages of star formation suggest that stars form in filaments, which results in a spatially substructured distribution of the stars that form \citep{Gomez93,Cartwright04,Sanchez09}. We impose substructure in both the spatial and velocity distributions  of the stars in our simulations by using the the \cite{goodwinDynamicalEvolutionFractal2004} box fractal method. 

We refer the interested reader to \citet{goodwinDynamicalEvolutionFractal2004} and \citet{DaffernPowell20} for full details of the method. We adopt a fractal dimension of $D = 2.0$, which results in a moderate degree of spatial and kinematic substructure and is commensurate with values measured in observed star-forming regions. 

In most of our simulations we create star-forming regions containing $N_\star = 100$ stars \citep[as these are unlikely to contain many, or any, massive stars that produce photoionising radiation that would preclude the formation of gas giant planets in our Solar System,][]{nicholsonSupernovaEnrichmentPlanetary2017,patelPhotoevaporationEnrichmentCradle2023a}. However, in one set of models we create simulations with $N_\star = 300$, $N_\star = 500$ and   $N_\star = 1000$ stars.  Individual stellar masses are sampled from the Maschberger IMF, described in the form of the following probability density function:
  
\begin{equation}
  P(M_\star) \propto \frac{M_\star}{\mu}^{-\alpha} \left( 1 + \left(\frac{M_\star}{\mu}\right)^{1-\alpha} \right)^{-\beta} ,
\end{equation}

\noindent
where $M_\star$ is the star mass, $\alpha$ is the high-mass exponent ($\alpha = 2.3$), $\beta$ is the low-mass exponent ($\beta = 1.4$), and $\mu$ is the scale parameter ($\mu = 0.2$) \citep{maschbergerFunctionDescribingStellar2013}.
Masses are constrained between $0.1 \, \si{\msol}$ and $50\,\si{\msol}$.
Stellar disks are assumed to have a radius of $400 \, \si{AU}$, similar to the radius assumption made in \cite{parkerIsotopicEnrichmentPlanetary2023}, based on the estimated average disk radii in star-forming regions proximal to NGC 2264 \citep{andrewsMassDependenceProtoplanetary2013,ciezaOphiuchusDIscSurvey2019}.
The lifetime of each disk is derived from an exponential probability density function with a mean lifetime of \SI{2}{Myr}; at the end of this lifetime, SLR mass fractions are stored.
Radioactive decay of SLRs within disks is modelled every time-step, and considered when determining the final SLR mass fraction after the disk has progressed to a protoplanetary system.
While stars could potentially form binary pairs the effect of binarity is not explicitly considered.

Much of the codebase of this project is shared with our previous paper, \citep{eatsonUnifiedInjectionModel2024}, which contains a more detailed explanation of all the shared routines with this code.

\subsection{Interloper enrichment}
\label{sec:int-enrich}

The main change between the simulation setup in this paper and the model described in \citet{eatsonUnifiedInjectionModel2024} is the addition of routines to calculate the amount of SLR enrichment in disks due to AGB interlopers.
Upon initialisation of the simulation, a single AGB star is created in addition to the star-forming region, at a distance of $x_i$ from the star-forming region and an initial velocity of $v_i$ heading towards the centre of the simulation (approximately the barycentre of the star-forming region).
A single AGB star is added to the $N$-body simulation, and thus can interact dynamically with the star-forming region. The AGB star's initial trajectory is towards the $(0,0,0)$ co-ordinate of the simulation at an initial speed of $v_i$.
Angled AGB trajectories are not simulated, instead we offset the AGB along the $y$ simulation axis with the parameter $y_i$.
Unless otherwise stated, the interloping star immediately enters the AGB phase at simulation start and begins depositing SLRs into the star-forming region, however the start time of the AGB phase can be offset with the parameter $\tau_i$.
As \seba{} does not support stars of mixed ages, stellar evolution of the AGB star itself is not considered, as such the mass of the interloping star may be higher than if the interloper were able to evolve, potentially influencing the dynamics of the star-forming region as the interloper passes through it.
Whilst this can result in differences in dynamics due to the changing mass of the AGB star, this was deemed an acceptable trade-off.

\begin{figure}
  \centering
  \includegraphics[width=\linewidth]{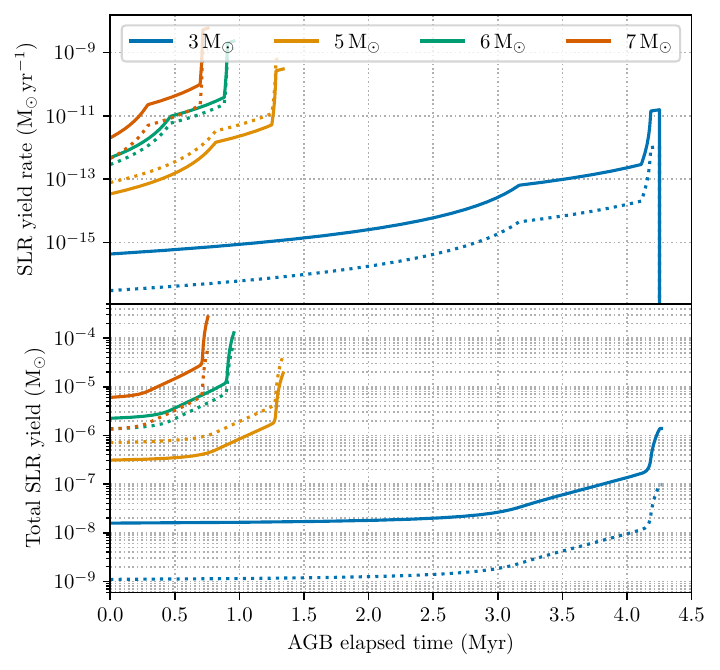}
  \caption{SLR yield rates and total yields for stars over their AGB phase, derived from the work in \citet{karakasStellarYieldsMetalRich2016}. Higher-mass stars have significantly shorter AGB phases, but have much higher emission rates and larger total yields.}
  \label{fig:agb-slr-yields}
\end{figure}

A series of lookup tables based on total SLR yields calculated in \cite{karakasStellarYieldsMetalRich2016} were formulated, the total yield was then spread over the lifetime of the AGB phase of the interloping star.
Lookup tables for \al{} and \fe{} were produced; the injection of other SLRs such as \atom{Ca}{41}, \atom{Mn}{53}, \atom{Pd}{107} and \atom{Hf}{182} were not modelled.
This assumes that the ratio of mass loss of SLRs from the interloper to the total wind mass loss is constant over the AGB phase, but this is more accurate than assuming that the SLR mass loss rate is constant and spread evenly over the AGB phase.
The total yield was used to calculate the SLR fraction of the total mass loss over the AGB phase, which was then used to estimate the SLR mass loss rate at the stars current age based on the AGBs current mass loss rate.
Fig. \ref{fig:agb-slr-yields} compares yield rates and total SLR yield for AGB stars of masses varying from $3\,\si{\msol}$ to $7\,\si{\msol}$, and shows the SLR mass loss rate steadily increases over time, and then rapidly surges at the end of the AGB phase.
\citet{parkerIsotopicEnrichmentPlanetary2023} details the results of changing interloper mass, while the higher-mass cases produce similar levels of enrichment -- $7\,\si{\msol}$ producing the most -- the low-mass $3\,\si{\msol}$ case produces insufficient SLR quantities (this can also be seen in Fig. \ref{fig:agb-slr-yields}).
For this paper we use a single discrete interloper mass case of $7\,\si{\msol}$.
By default, the interloper enters the AGB star immediately after the start of the simulation, but this can be offset at runtime, resulting in no enrichment until a defined time; this is referred to in our work as the interloper offset time, $\tau_i$.

Disk SLR enrichment is calculated using the method described in \citet{eatsonUnifiedInjectionModel2024}, though with the aforementioned AGB enrichment calculated as well.
Similar to the model described in \citet{parkerIsotopicEnrichmentPlanetary2023}, we incorporate a cylindrical model of outflow from the interloper, as the flow is substantially slower than the Wolf-Rayet winds simulated in our previous paper.
However, to account for deflection and potential trapping of the AGB by the star-forming region, we model the outflow region as a series of connected cylindrical segments.
As the interloper moves through the simulation a new one is added every simulation timestep for the lifetime of the star, SLRs are deposited into the segment based on the current AGB SLR mass loss rate, $\dot{\mass}\rms{SLR}$, such that:

\begin{equation}
  \mass\rms{SLR,seg} = \dot{\mass}\rms{SLR} \Delta t,
\end{equation}

where $\Delta t$ is the simulation timestep.
The segment has a length equivalent to the distance travelled by the interloper between each timestep and a radius of $0.1\,\si{pc}$.
After the $N$-body simulation has evolved at each timestep, every remaining disk is checked to see whether it is currently inside or has passed through a segment, with the fraction of the timestep that the star has intersected a cylinder segment defined as $\eta\rms{int}$.

The total mass sweep-up efficiency of the disk is calculated with the equation:

\begin{equation}
  \eta\rms{sw} = \frac{r\rms{disk}^2}{r\rms{cyl}^2} \frac{d\rms{trav}}{l\rms{cyl}} ,
\end{equation}

\noindent
where $r\rms{disk}$ is the disk radius, $d\rms{trav}$ is the distance travelled within the cylinder by the approaching star during a timestep and $r\rms{cyl}$ is the cylinder radius.
This is equivalent to the ratio of the volumes of two cylinders, the previously described segment and the straight-line path of the intersecting disk.
The total disk sweep-up mass ($\mass\rms{SLR,sw}$) can be then calculated with the formula:

\begin{equation}
  \mass\rms{SLR,sw} = \eta\rms{sw} \eta\rms{int} \mass\rms{SLR,seg} .
\end{equation}

\noindent
The amount of material removed from the segment is removed from the total SLR mass reserved of the segment, available SLR mass is reduced every time-step to account for radioactive decay processes.
For a given simulation of a timestep of $10^4 \,\si{yr}$, this results in $\approx 70$ discrete segments for a $7\,\msol$ star. Segments are not dispersed over time, as the dispersion time of this wind-blown region could vary significantly and would introduce a series of free parameters into the simulation; we therefore assume that the dispersion time is longer than the AGB timescale of the interloping star.

This method expands on work by \cite{parkerIsotopicEnrichmentPlanetary2023} by considering the trajectory of the AGB star, as well as a more accurate estimation of how disks interact with the AGB star via the changing trajectory of the AGB rather than a static cylinder.
Future research could involve replacing the segmented cylinder with an SPH wind model, where density can be accurately sampled and effect of the AGB interloper moving through the local medium could be better simulated.
Expansion of the cylinders over time to account for dispersion could also be performed.

The properties of the interloper can be varied at simulation start time, such parameters include the interloper mass ($\mass_i$), the interlopers initial velocity towards the cluster ($v_i$) and the interlopers initial position relative to the centre of the star-forming region along the $x$ and $y$ simulation axes ($x_i$ and $y_i$).
As with the previous paper \citep{eatsonUnifiedInjectionModel2024}, the radius  of the star-forming region $r_c$, and total region population size, $N_\star$, can also be varied.

\subsection{Parameter space explorations}

\begin{table*}
\centering
\caption{A summary of the model sets performed in this paper, detailing which parameters are varied, by how much, whether simulations are repeated, as well as brief notes about the specifics of the simulation set are also included.}
\label{tab:properties}
\begin{tabular}{cccccccc}
\hline
 Set & Varied Parameters & $r_c$ range (pc) & $v_i$ range (km\,s$^{-1}$) & $\tau_i$ range (Myr) & $N_\star$ range & Repetition & Notes \\ \hline
Set A & $r_c$, $v_i$ & 0.1, 0.3, 1.0 & 1, 3, 10, 30 & 0.0 & 100 & Yes & Broad parameter search \\
Set B & $\tau_i$, $v_i$ & 0.3 & 0.1-100 & 0.01-10 & 100 & None & Logarithmically spaced search \\
Set C & $x_i$, $y_i$, $v_i$ & 0.3 & 1, 3, 10, 30 & 0 & 100 & None & \SI{8}{pc} $x_i$ range, \SI{1}{pc} $y_i$ range \\
Set C 2 & $x_i$, $v_i$ & 0.3 & 0 & 0 & 100 & None & \SI{25}{pc} $x_i$ range \\
Set D & $N_\star$, $v_i$ & 0.3 & 0 & 0 & 100, 300, 500, 1000 & Varies by $N_\star$ & Changes population size
\end{tabular}%
\end{table*}

The simulations performed in this paper can be broadly divided into multiple sets:

\begin{itemize}
  \item Set A: A broad parameter space search varying region radius, $r_c$, and interloper velocity, $v_i$.
  \item Set B: A series of simulations varying $v_i$ and interloper AGB phase start time, $\tau_i$.
  \item Set C: A series of simulations varying AGB initial position ($x_i,y_i$), as well as $v_i$.
  \item Set D: A series of simulations with larger stellar populations.
\end{itemize}

In order to ascertain which parameters of the star-forming region and interloper are influential, simulation set A is performed to determine the influence of the system parameters, as well as to test the interloper injection routines.
Parameter combinations are repeated to rule out variation due to the structure and composition of the randomly generated clusters.

Once the ideal parameters for the interloper mass ($7\,\text{M}_\odot$) and region radius ($0.3\,\si{pc}$) were established from the results of set A, we progressed to a finer parameter space search that varied $v_i$ and $\tau_i$. No repetition was performed on this set; instead $\tau_i$ was varied from $10^{-2}\,\si{Myr}$ to $10\,\si{Myr}$ and $v_i$ was varied from $0.1\,\si{km.s^{-1}}$ to $100\,\si{km.s^{-1}}$ in 32 logarithmically spaced steps each, leading to a total of 1024 discrete simulations.
These simulations establish how influential the initial velocity and condition of the interloper is to the enrichment of a star-forming region.

Set C was designed to determine the effect of ``near-miss'' interactions between the interloper and the star-forming region.
The initial position of the interloper ($x_i$,$y_i$) was varied between $-8\,\si{pc}$ and $0\,\si{pc}$ in the $x$ simulation axis and $0\,\si{pc}$ to $1\,\si{pc}$ in the $y$ simulation axis, where $(0,0$) is the initial centre of the star-forming region.
The interloper immediately enters the AGB phase, unlike set B, and has an initial velocity of $1\,\si{km.s^{-1}}$, $3\,\si{km.s^{-1}}$, $10\,\si{km.s^{-1}}$ or $30\,\si{km.s^{-1}}$ on the $x$ axis towards the star-forming region.
$x$ and $y$ axis stepping is evenly spaced, with $64$ discrete positions on the $x$ axis and $8$ discrete positions on the $y$ axis, and similarly to set B, there is no repetition.
Another sub-set of set C is also performed, where $y_i$ is not varied, but the $x_i$ range is increased to $25\,\si{pc}$, in order to observe enrichment due to more evolved, high-velocity interlopers.
Whilst all previous simulation sets have a region population of $N_\star = 100$, set D differs by varying the population to either $N_\star = 300$, $N_\star = 500$ or $N_\star = 1000$.

Based on our results from set C the interloper for all simulations had an initial distance from the star-forming region of $x_i = 3.0\,\si{pc}$.
Simulation repetition was performed, with larger simulations having fewer repetitions in order for each parameter space set to have the same number of stars simulated, and to accommodate for increased computational complexity due to the increased number of stars.
The interloper radius was also varied from $1\,\si{km.s^{-1}}$ to $30\,\si{km.s^{-1}}$.
An increased population will increase the total mass of the star-forming region, resulting in a greater escape velocity. This could enhance the likelihood of trapping lower velocity AGB stars.

\section{Results}
\label{sec:results}

\subsection{Set A: Initial simulations}
\label{sec:seta}
\begin{table*}
  \centering
  \caption{A collection of results from simulation set A, with the initial simulation parameters detailed. Uncertainties are based on running simulations with identical parameters 50 times. We show the simulation subset, radius of the star-forming region $r_c$, velocity of the interloper $v_i$, the number of disks summed across all realisations of these simulations, $N\rms{disk,tot}$. We then show the number of disks that have any enrichment, $Z\rms{en}$, the number of disks enriched to a tenth of the \al{} in the Solar system $Z_\mathrm{26Al,0.1SS}$, and the measured \al{} in the Solar system, $Z_\mathrm{26Al,SS}$, the number of disks enriched to the lower estimate for \fe{} in the Solar system,  $Z_\mathrm{60Fe,Lo}$, the number enriched to the higher estimate for   \fe{} in the Solar system,  $Z_\mathrm{60Fe,Hi}$ and the mass of \al{} and \fe{} injected into the disks, $\mass_\mathrm{26Al,inj}$ and $\mass_\mathrm{60Fe,inj}$ respectively. Enrichment is very dependent on the original velocity of the interloper, $v_i$, though even in the case of simulations with an interloper velocity on the order of $10\,\si{km.s^{-1}}$ we observe a few disks obtaining \al{} enrichment levels sufficient for volatile desiccation in planetesimals \citep{eatsonDevolatilizationExtrasolarPlanetesimals2024,lichtenbergWaterBudgetDichotomy2019}. Similarly, \fe{} enrichment to the low Solar estimate is surprisingly common for faster interloper enrichment. However, interlopers moving at or above the galactic velocity dispersion ($\approx 30\,\si{km.s^{-1}}$) are not conducive to enrichment, at least with an initial position of 1 parsec from the centre of the star-forming region.}
  \resizebox{\textwidth}{!}{%
  \begin{tabular}{ccccccccccc}
  \hline
  Simulation Subset & $r_c$ & $v_i$ & $N\rms{disk,tot}$ & $Z\rms{en}$ & $Z_\mathrm{26Al,0.1SS}$ & $Z_\mathrm{26Al,SS}$ & $Z_\mathrm{60Fe,Lo}$ & $Z_\mathrm{60Fe,Hi}$ & $\mass_\mathrm{26Al,inj}$ & $\mass_\mathrm{60Fe,inj}$ \\
   & pc  & \si{km.s^{-1}} & &  &  &  & &  & $\msol$ & $\msol$ \\
  \hline
\texttt{set-a-rc-0.1-inv-1.00} & 0.1 & 1.0 & 4841 & $0.29 \pm 0.04$ & $0.27 \pm 0.04$ & $0.24 \pm 0.03$ & $0.29 \pm 0.04$ & $0.29 \pm 0.04$ & $\left(2.8 \pm 0.7\right) \times 10^{-6}$ & $\left(8.7 \pm 2.2\right) \times 10^{-7}$ \\
\texttt{set-a-rc-0.1-inv-3.00} & 0.1 & 3.0 & 4840 & $0.64 \pm 0.03$ & $0.46 \pm 0.03$ & $0.23 \pm 0.02$ & $0.63 \pm 0.03$ & $0.39 \pm 0.03$ & $\left(3.66 \pm 0.35\right) \times 10^{-8}$ & $\left(1.17 \pm 0.13\right) \times 10^{-8}$ \\
\texttt{set-a-rc-0.1-inv-10.0} & 0.1 & 10.0 & 4844 & $0.81 \pm 0.01$ & $0.17 \pm 0.01$ & $0.0002 \pm 0.0002$ & $0.70 \pm 0.02$ & $0.0004 \pm 0.0003$ & $\left(2.10 \pm 0.13\right) \times 10^{-9}$ & $\left(6.0 \pm 0.4\right) \times 10^{-10}$ \\
\texttt{set-a-rc-0.1-inv-30.0} & 0.1 & 30.0 & 4872 & $0.890 \pm 0.008$ & $0.008 \pm 0.002$ & $0.0$ & $0.49 \pm 0.02$ & $0.0$ & $\left(4.88 \pm 0.22\right) \times 10^{-10}$ & $\left(1.30 \pm 0.07\right) \times 10^{-10}$ \\
\texttt{set-a-rc-0.3-inv-1.00} & 0.3 & 1.0 & 4839 & $0.17 \pm 0.02$ & $0.14 \pm 0.01$ & $0.11 \pm 0.01$ & $0.17 \pm 0.02$ & $0.16 \pm 0.01$ & $\left(5.7 \pm 0.7\right) \times 10^{-7}$ & $\left(1.90 \pm 0.26\right) \times 10^{-7}$ \\
\texttt{set-a-rc-0.3-inv-3.00} & 0.3 & 3.0 & 4847 & $0.37 \pm 0.02$ & $0.22 \pm 0.01$ & $0.075 \pm 0.005$ & $0.37 \pm 0.02$ & $0.152 \pm 0.009$ & $\left(1.05 \pm 0.07\right) \times 10^{-8}$ & $\left(3.8 \pm 0.4\right) \times 10^{-9}$ \\
\texttt{set-a-rc-0.3-inv-10.0} & 0.3 & 10.0 & 4867 & $0.47 \pm 0.01$ & $0.047 \pm 0.003$ & $0.0$ & $0.38 \pm 0.01$ & $0.0002 \pm 0.0002$ & $\left(6.69 \pm 0.23\right) \times 10^{-10}$ & $\left(2.17 \pm 0.09\right) \times 10^{-10}$ \\
\texttt{set-a-rc-0.3-inv-30.0} & 0.3 & 30.0 & 4855 & $0.48 \pm 0.01$ & $0.0004 \pm 0.0003$ & $0.0$ & $0.188 \pm 0.007$ & $0.0$ & $\left(1.53 \pm 0.05\right) \times 10^{-10}$ & $\left(4.56 \pm 0.14\right) \times 10^{-11}$ \\
\texttt{set-a-rc-1.0-inv-1.00} & 1.0 & 1.0 & 4846 & $0.026 \pm 0.003$ & $0.019 \pm 0.002$ & $0.011 \pm 0.002$ & $0.026 \pm 0.003$ & $0.021 \pm 0.002$ & $\left(3.9 \pm 1.0\right) \times 10^{-8}$ & $\left(1.41 \pm 0.31\right) \times 10^{-8}$ \\
\texttt{set-a-rc-1.0-inv-3.00} & 1.0 & 3.0 & 4845 & $0.081 \pm 0.004$ & $0.039 \pm 0.003$ & $0.014 \pm 0.002$ & $0.078 \pm 0.004$ & $0.029 \pm 0.002$ & $\left(2.1 \pm 0.4\right) \times 10^{-9}$ & $\left(1.07 \pm 0.34\right) \times 10^{-9}$ \\
\texttt{set-a-rc-1.0-inv-10.0} & 1.0 & 10.0 & 4838 & $0.096 \pm 0.005$ & $0.006 \pm 0.001$ & $0.0$ & $0.075 \pm 0.004$ & $0.0$ & $\left(9.5 \pm 0.6\right) \times 10^{-11}$ & $\left(3.58 \pm 0.22\right) \times 10^{-11}$ \\
\texttt{set-a-rc-1.0-inv-30.0} & 1.0 & 30.0 & 4862 & $0.074 \pm 0.004$ & $0.0$ & $0.0$ & $0.025 \pm 0.002$ & $0.0$ & $\left(1.56 \pm 0.09\right) \times 10^{-11}$ & $\left(5.70 \pm 0.35\right) \times 10^{-12}$

  \end{tabular}%
  }
  \label{tab:set-a-results}
  \end{table*}

\begin{figure*}
  \centering
  \includegraphics[scale=0.7]{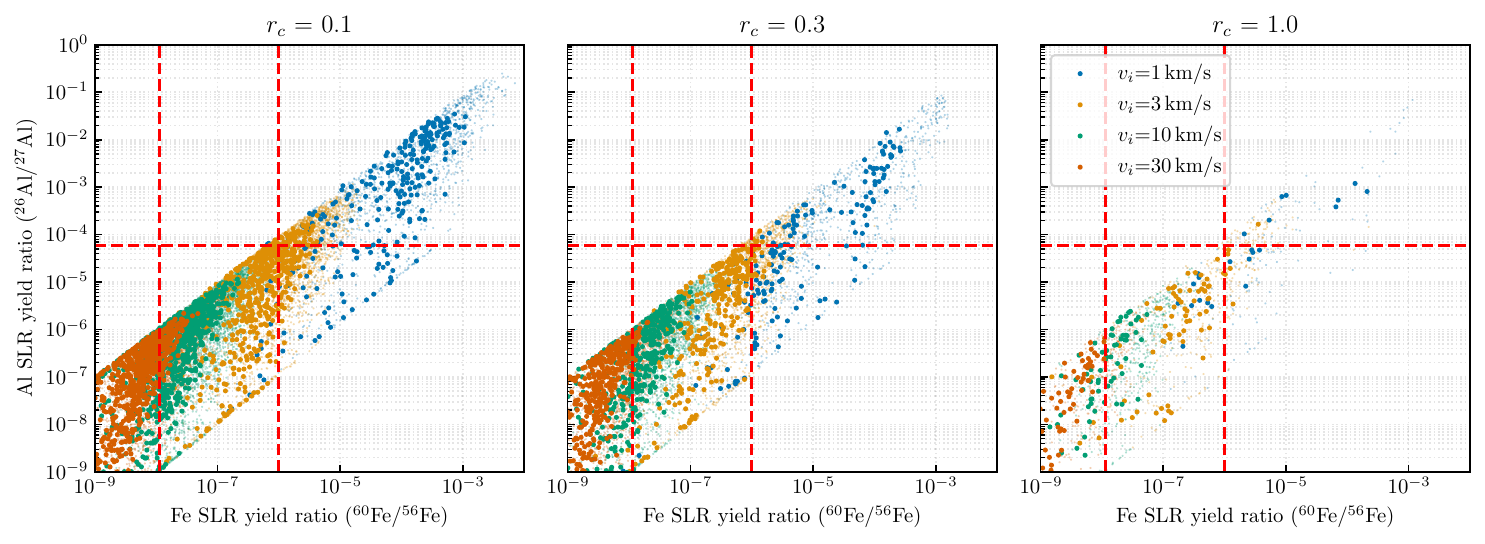}
  \caption{A comparison of \al{} and \fe{} enrichment for set A simulations, split by $r_c$. Opaque, larger markers represent disks with a host star between \SI{0.5}{\msol} and \SI{1.5}{\msol}. Disks with Solar System levels with a Solar-like host star are possible, though infrequent. Dashed lines represent estimated Solar System formation levels of enrichment based on observational data. Enrichment amount does not vary considerably as $r_c$ is varied, though the probability of the disks undergoing high levels of enrichment decreases significantly. There is a clear inverse relationship between $v_i$ and enrichment of either isotope.}
  \label{fig:set-a-isotope}
\end{figure*}

For this simulation set, star-forming region radius, $R_c$, interloper mass, $\mass_i$, and interloper initial velocity, $v_i$, are varied in wide steps.
Star-forming regions have $N_\star = 100$ stars, an initial radius of 0.1, 0.3 or 1.0 parsecs, which corresponds to a stellar density ($\tilde{\rho}$) of $\sim \num{2e4}\,\si{pc^{-3}}$, $\sim 10^3\,\si{pc^{-3}}$ and $\sim 20\,\si{pc^{-3}}$ respectively.
Interlopers are introduced into the simulation with a velocity of either 1, 3, 10 or 30 \si{km.s^{-1}} into a star-forming region of radius 0.1, 0.3 or 1.0 pc.
The initial position of the interloper was 1 parsec from the simulation centre, which is the approximate centre of the star-forming region as well.
Parameter combinations were repeated 50 times to rule out random effects and variation due to initial cluster morphology, this is also used to calculate uncertainty for our results.
This set acts as an initial parameter space exploration using our initial model, and is also used to verify whether the model is accurate by comparing it to previous work by \citet{parkerIsotopicEnrichmentPlanetary2023}.

Two useful values we will consider in our assessment are the \al{} significant enrichment fraction, $Z\rms{26Al,0.1SS}$, and the Solar System \al{} enrichment fraction, $Z\rms{26Al,SS}$ \citep{thraneExtremelyBriefFormation2006}, which is the fraction of disks that undergo enrichment sufficient to affect planetesimal evolution, and the fraction of disks that undergo enrichment equivalent to the Solar System estimate for a given SLR.
We define $Z\rms{26Al,0.1SS}$ as a tenth of the early-Solar \al{} enrichment fraction due to simulations by \citet{lichtenbergWaterBudgetDichotomy2019} as well as our previous results on planetesimal simulation \citep{eatsonDevolatilizationExtrasolarPlanetesimals2024}.
It was found that at around this amount, out-gassing began to occur in a \SI{100}{km} planetesimal.
The effects of \fe{} heating are less pronounced, with significant outgassing due to \fe{} occurring at $Z\rms{60Fe}$ values that are two orders of magnitude higher than the higher Solar System enrichment estimate of \fe{} \citep{eatsonDevolatilizationExtrasolarPlanetesimals2024}.
We consider two different values for quantifying the number of disks that have significant $Z\rms{60Fe}$ enrichment; the  fraction of disks that exceed the low \citep[$Z\rms{60Fe} > 10^{-8}$,][]{tangAbundanceDistributionOrigin2012}  and high \citep[$Z\rms{60Fe} > 10^{-6}$,][]{mishraAbundance60FeInferred2016}  estimates for Solar System \fe{} enrichment \citep[see also][]{Cook21,Kodolanyi22}. These are labelled as $Z\rms{60Fe,Lo}$ and $Z\rms{60Fe,Hi}$ in Table~\ref{tab:set-a-results}, respectively.
We also record the total mass of \al{} and \fe{} injected into disks in the star-forming region, $\mass\rms{SLR,inj}$, which we can compare with the total SLR mass loss of a \SI{7}{\msol} star.
We calculate these total mass losses using the SLR yields in \citet{karakasStellarYieldsMetalRich2016}, and were found to be $\mass\rms{26Al} \approx \SI{2.7e-4}{\msol}$ and $\mass\rms{60Fe} \approx \SI{6.0e-05}{\msol}$.

Table \ref{tab:set-a-results} contains the results for each combination of parameters performed in set A.
We find that there is a wide variance of enrichment levels for disks, but there is a strong clustering correlation by parameter sub-set, as can be clearly seen in Fig. \ref{fig:set-a-isotope}.
With AGB enrichment disks can be enriched to values substantially exceeding Solar System levels, though the extremely highly enriched examples may not be possible, as we will later discuss.
However, for more realistic parameters, early-Solar levels of enrichment are indeed possible. 
We find that the cluster radius does not have an effect on the maximum amount of enrichment for a simulation subset, though it can drastically effect the number of stars undergoing enrichment, resulting in a total decrease in the fraction of disks enriched to any amount, $Z\rms{en}$, of approximately an order of magnitude between the \SI{0.1}{pc} and \SI{1.0}{pc}.
Influential enrichment levels are possible for extremely sparse star-forming regions, though this is much less likely.
For successive simulation sets we adopted $r\rms{c} = \SI{0.3}{pc}$, similar to much of our previous work.
The most important parameter explored was the initial velocity of the interloper, as the velocity increases $Z\rms{en}$ increases, though there is a marked decrease in the number of stars that undergo significant amounts of enrichment as the velocity increases above \SI{3}{\kms}, \SI{1}{\kms} cases have a reduced $Z\rms{en}$.
While counter-intuitive, this is due to taking $\sim \SI{1}{Myr}$ to reach the centre of the simulation, and thus would only interact with the outlying stars before ending the AGB phase.
This elapsed time would also reduce the number of viable disks the AGB wind could interact with.
In terms of the total SLR mass injected into every disk in the simulation, the slower simulations were found to inject far more material into the disks, with an efficiency of about 1\%, this decreased substantially as interloper velocity and $r\rms{c}$ increase, though this still results in enrichment to levels that can influence the thermo-chemical evolution of subsequent planetesimals.

Compared to the Galactic velocity dispersion (\SI{30}{\kms}), \SI{3}{\kms} to \SI{10}{\kms} would constitute a relatively slow, ``walk-away'' velocity star \citep{Schoettler20,Schoettler22}, but is within a reasonable value and coincides with the relative velocities of stars near to the Solar System. The \SI{1}{\kms} case, or stars lower than that, would be extremely unlikely.
As such, our initial estimates suggest that AGB enrichment is a viable method of disk SLR enrichment, our subsequent simulation sets determine how sensitive this amount of enrichment is to the interlopers properties, as well as the initial conditions of the star-forming region.
It is also important to note that while \al{} is a more important source of radioisotopic heating in the early Solar System, the interloper enrichment mechanism can produce \fe{} enrichment in disks equivalent to early-Solar System estimates.
This level of  \fe{} enrichment was determined in our previous work to not be possible with Wolf-Rayet winds \citep{eatsonUnifiedInjectionModel2024}, and would require supernovae, which would destroy the gas within the disks \citep{patelPhotoevaporationEnrichmentCradle2023a}, resulting in a lack of gas giant planets in the planetary system(s).

\subsection{Set B: AGB time offset}
\label{sec:setb}

\begin{figure}
  \centering
  \includegraphics[width=\linewidth]{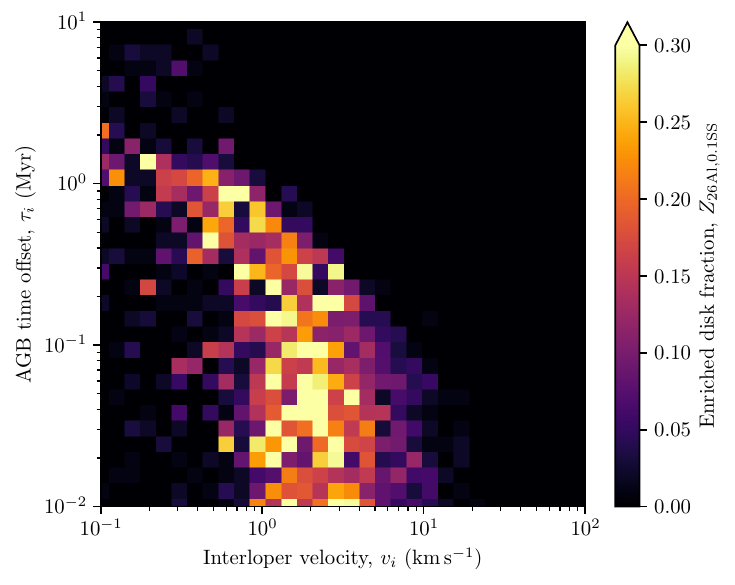}
  \caption{Enrichment fractions for each simulation in the B set. There is a strong dependency on interloper velocity and yield, with a significant reduction in enrichment as the initial interloper velocity exceeds $10\,\si{km.s^{-1}}$. Whilst there is also an inverse relationship with enrichment and interloper AGB phase offset time, this is largely due to a lack of viable protoplanetary disks.}
  \label{fig:set-b-enrichment}
\end{figure}

\begin{figure}
  \centering
  \includegraphics[width=\linewidth]{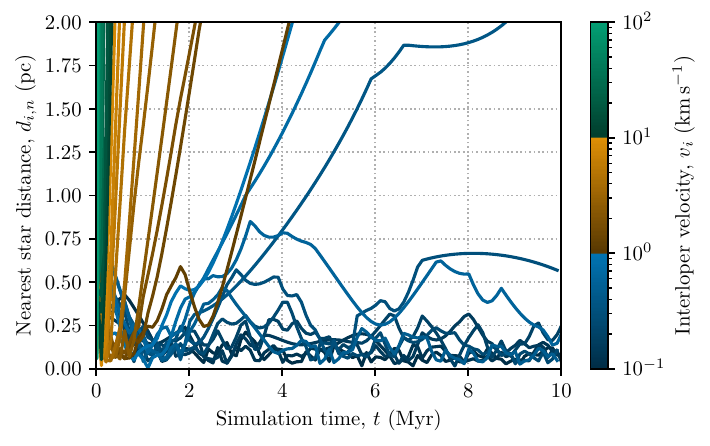}
  \caption{Interloper distance from the nearest star over time for simulations where the interloper enters the AGB phase immediately after the start of the simulation. Most interlopers do not become trapped in the star-forming region.}
  \label{fig:set-bdistance-from-centre}
\end{figure}

In this set of simulations, the variable parameters are the initial interloper velocity, $v_i$ and the interloper time offset, $\tau_i$.
$v_i$ is logarithmically spaced varying from $0.1\,\si{\kms}$ to \SI{100}{\kms}, $\tau_i$ was similarly spaced between \SI{0.01}{Myr} to \SI{10}{Myr}.
The interloper was placed at a distance of \SI{1}{pc} from the centre of the star-forming region, and the region has a radius of $r\rms{c}=\SI{0.3}{pc}$.
Fig. \ref{fig:set-b-enrichment} shows the enrichment fractions for each simulation in this set, in ideal conditions the number of disks in a simulation that undergo near-Solar enrichment can exceed 30\%, with some consistency.
Whilst this puts the solar system in the upper limit of disk enrichment if our results in set A are also considered, this suggests that if AGBs encounter disks at ``walk-away'' speeds -- below the velocity dispersion of the Milky Way -- \al{} enrichment would be an important driving force in planetesimal formation \citep{lichtenbergWaterBudgetDichotomy2019,eatsonDevolatilizationExtrasolarPlanetesimals2024}.
Broadly speaking, significant disk enrichment becomes common between \SI{1}{\kms} and \SI{10}{\kms}, with a relatively sharp cut-off above \SI{10}{\kms}.
While the high velocity of the interloper is a contributing factor to this lack of enriched disks -- as SLRs are spread more sparsely over a wider area -- another reason would be that the interloper simply passes by the cluster before AGB outflow is heightened, a few hundred thousand years into its lifespan.
This can be observed in the lower-velocity simulations with a higher value for $\tau_i$, and is investigated further in Set C by varying the interlopers initial distance to the cluster instead.
Furthermore, as the disk population halves every \SI{2}{Myr}, this can also reduce the enriched disk population.
Fig. \ref{fig:set-bdistance-from-centre} shows distances from AGB interlopers to their nearest cluster stars for a sample of the set B data where $\tau_i = \SI{0.01}{Myr}$. In the case of very low velocity interlopers, the interloper does not continue its progress away from the star-forming region, instead becoming ``trapped''.
This is to be expected, as these velocities are below the regions escape velocity, however, this leads to the interloper interacting with nearly every disk in the region, and would also lead to multiple interaction occurrences for individual disks.
The probability of this occurring is potentially low, however, but would lead to extremely enriched disks in star-forming regions.
Such a study is beyond the scope of this paper, but would deem investigation if ``trapped'' interlopers were found in observed regions.

\subsection{Set C: AGB distance offset}
\label{sec:setc}

\begin{figure*}
  \centering
  \includegraphics[scale=0.7]{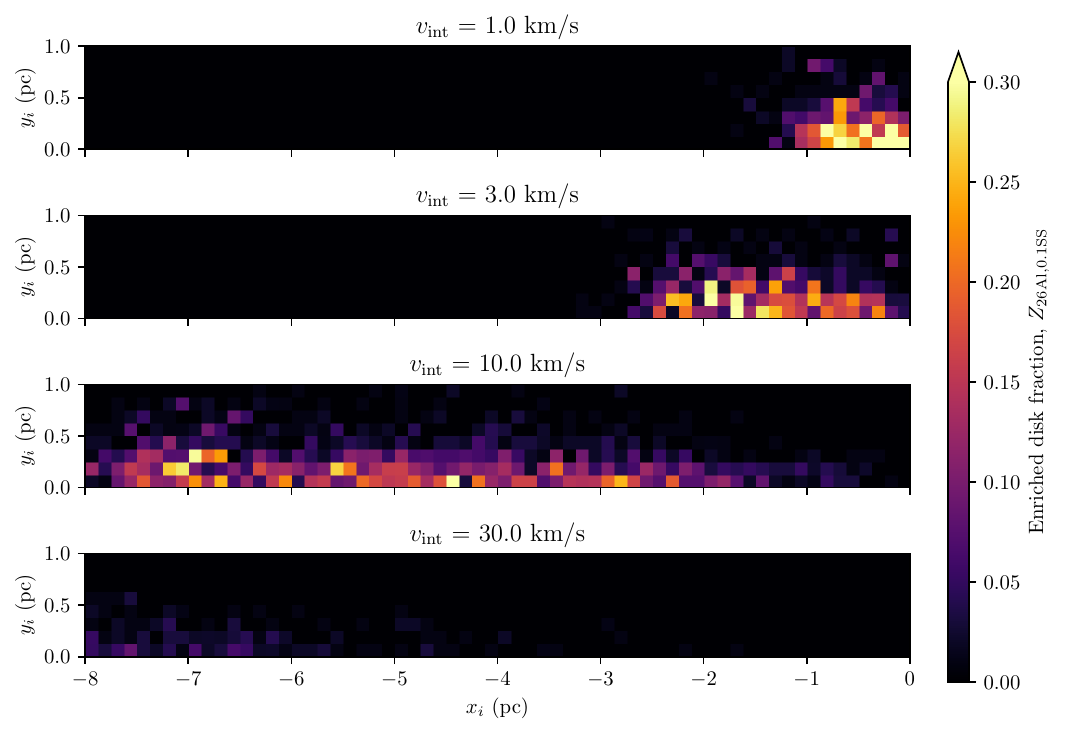}
  \caption{Comparison of enrichment in simulations where interloper $x_i$ and $y_i$ offset are changed. Results are separated by initial interloper velocity. The $10\,\si{km.s^{-1}}$ case consistently produces some highly enriched disks, as long as the interloper is given some time to evolve before it interacts with the star-forming region.}
  \label{fig:set-c-pos-plot}
\end{figure*}

\begin{figure}
  \centering
  \includegraphics[scale=0.7]{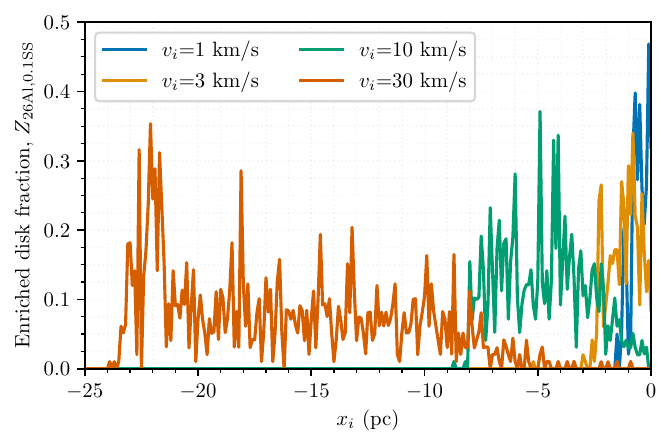}
  \caption{Comparison of \al{} enrichment in simulations where $x_i$ and $v_i$ are changed. An expanded and higher density $x_i$ parameter space is used compared to the data in Fig. \ref{fig:set-c-pos-plot}, we find that lowered, but still viable levels of Solar-like \al{} enrichment occur in the \SI{30}{\kms} case, suggesting that main influences on enrichment are the progression of the interloper into the AGB phase as well as the interlopers velocity.}
  \label{fig:set-c-expanded}
\end{figure}

\begin{figure}
  \centering
  \includegraphics[scale=0.7]{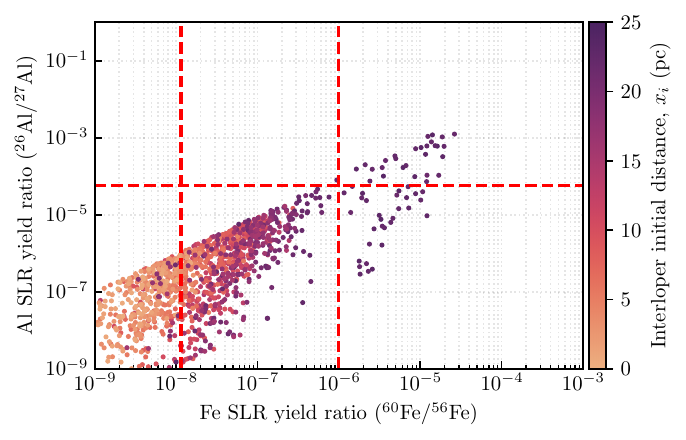}
  \caption{A comparison of isotopic ratios for all disks with Solar-like parent stars in the Fig. \ref{fig:set-c-expanded} subset where $v_i = \SI{30}{\kms}$. A small number of disks have significantly elevated SLR enrichment, to the point where even \fe{} would be an important desiccant, though \al{} would still be the dominant heating source in eventual planetesimal evolution.}
  \label{fig:set-c-expanded-isotope}
\end{figure}

In order to verify where significant enrichment of disks is possible via higher-velocity interlopers, in this set we start the interloper at a varying distance to the star-forming region, from \SI{-8}{pc} to \SI{0}{pc}.
We also vary the $y$ axis offset from \SI{0}{pc} to \SI{1}{pc} -- or 3 times the region radius -- to determine the effect on enrichment due to ``near misses'', where the interloper does not directly run through the centre of mass of the star-forming region.
We also vary the velocity of the AGB from \SI{1}{\kms} to \SI{30}{\kms}.
Fig. \ref{fig:set-c-pos-plot} clearly shows again that \SI{10}{\kms} is a very viable velocity for disk enrichment, and enrichment is not significantly influenced by initial distance, so long as the interloper encounters the cluster after it has had time to reach a point where SLR production is increased compared to the start of the AGB phase.
In general, in fact, velocity increases, the parameter space where disk enrichment is viable spreads out.
The average fraction of suitably enriched disks does also decrease as velocity increases, due to a sparser field of SLRs for the disks to sweep up.
However, by \SI{30}{\kms} disk enrichment to sufficient levels for thermo-chemical evolution does still occur, just in numbers.
This is due to the resulting cloud of SLRs from the wind outflow being too diffuse, and as such sweep-up per unit time becomes too low.

Fig. \ref{fig:set-c-expanded} contains data from a sub-set of set C where $y_i$ was not varied, but $x_i$ was varied out to \SI{25}{pc} instead, with steps every \SI{0.1}{pc}.
This was performed in order to better determine the variability between simulations and to also determine if \SI{30}{\kms} cases could produce a large population of disks.
We found that \SI{30}{\kms} could in fact produce lower, but still prominent ($\sim 10\%$) fractions of enriched disks.
This enrichment fraction as a function of distance is similarly flat beyond a minimum value like the \SI{10}{\kms} case in Fig. \ref{fig:set-c-pos-plot}, though with a more prominent peak.
Filtering the subset down to disks with Solar-like host stars and comparing isotopic enrichment instead (Fig. \ref{fig:set-c-expanded-isotope}), we find that very high levels of enrichment can occur, to the point where \fe{} enrichment could be influential in devolatilization \citep{eatsonDevolatilizationExtrasolarPlanetesimals2024}.
The results of set B and C suggest that the interloper being in the latter part of its AGB phase is extremely important to disk enrichment, with interlopers with velocities at or above the galactic velocity dispersion ($\sim\SI{30}{\kms}$) able to enrich disks to levels where substantial desiccation can occur.

\subsection{Set D: larger simulations}
\label{sec:setd}

\begin{figure*}
  \centering
  \includegraphics[scale=0.7]{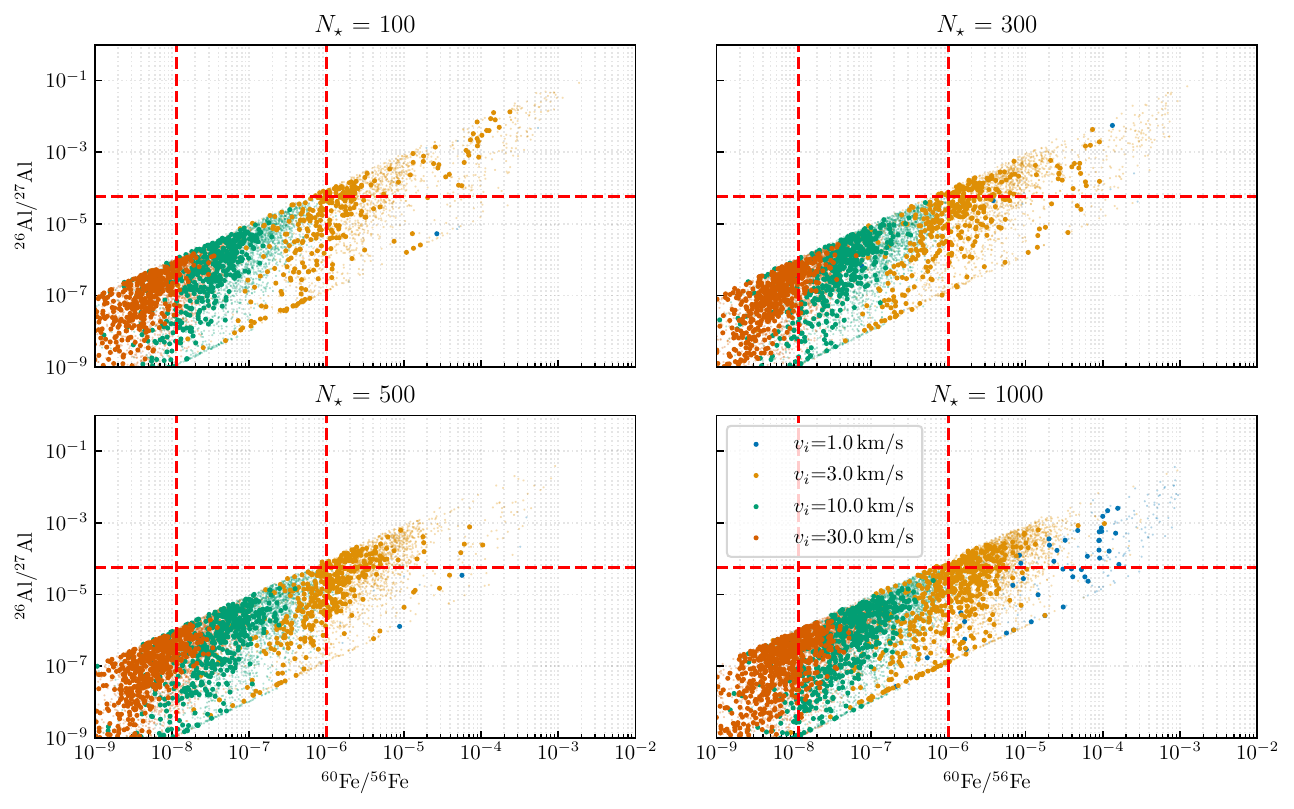}
  \caption{A comparison of \al{} and \fe{} enrichment for set D simulations, split by $N_\star$. We see that there is a minor improvement in enrichment in larger star-forming regions in general, however, enrichment is still very velocity dependent, and high-velocity \SI{30}{km.s^{-1}} interlopers do not enrich to Solar System levels.}
  \label{fig:setd-isotopes}
\end{figure*}

\begin{figure*}
  \centering
  \includegraphics[scale=0.7]{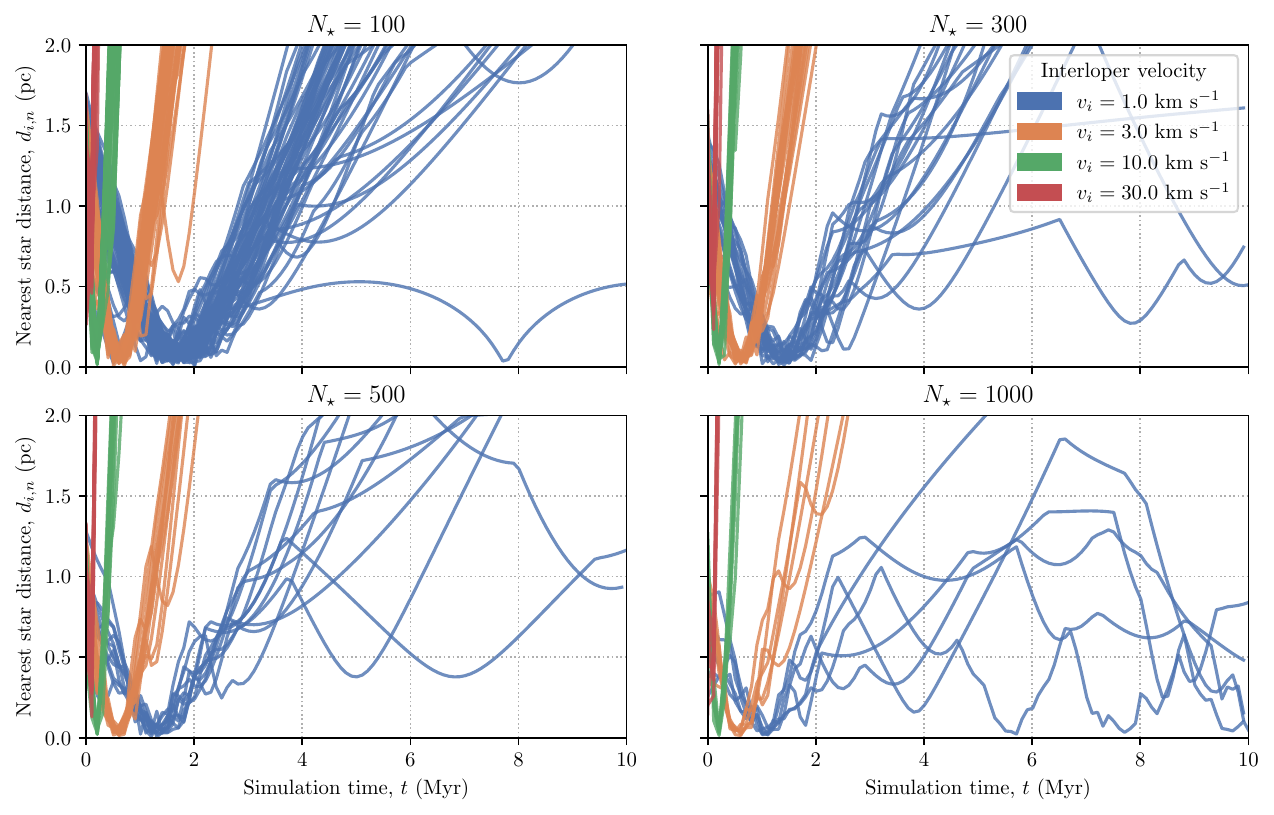}
  \caption{A comparison of the distance between the interloper and the nearest star for the simulations in set D over time, separated by interloper velocity, $v_i$. This is a similar plot to Fig. \ref{fig:set-bdistance-from-centre}.}
  \label{fig:setd-com}
\end{figure*}

\begin{table*}
  \centering
  \caption{A table of the averaged results for each sub-set of the simulations in set D, similar to Table \ref{tab:set-a-results}. There is a small correlation on Solar System level enrichment probability and star count, however, this is relatively minor.}
  \label{tab:setd}
  \resizebox{\textwidth}{!}{%
  \begin{tabular}{ccccccccccc}
  \hline
  Simulation Name & $N_\star$ & $v_i$ & $N_{\text{disk,tot}}$ & $Z\rms{en}$ & $Z_\mathrm{26Al,0.1SS}$ & $ Z_\mathrm{26Al,SS}$ & $Z_\mathrm{60Fe,Lo}$ & $Z_\mathrm{60Fe,Hi}$ & $\mass_\mathrm{26Al,inj}$ & $\mass_\mathrm{60Fe,inj}$ \\
  && \si{km.s^{-1}} &&&&&&&$\msol$ &$\msol$  \\ \hline
\texttt{set-d-v-1.00-ns-0100} & 100 & 1.0 & 5857 & $0.0012 \pm 0.0004$ & $0.0008 \pm 0.0004$ & $0.0005 \pm 0.0003$ & $0.0012 \pm 0.0004$ & $0.0012 \pm 0.0004$ & $\left(1.2 \pm 0.9\right) \times 10^{-9}$ & $\left(9 \pm 5\right) \times 10^{-10}$ \\
\texttt{set-d-v-3.00-ns-0100} & 100 & 3.0 & 5825 & $0.26 \pm 0.01$ & $0.194 \pm 0.009$ & $0.120 \pm 0.008$ & $0.26 \pm 0.01$ & $0.20 \pm 0.01$ & $\left(1.66 \pm 0.31\right) \times 10^{-7}$ & $\left(5.9 \pm 1.0\right) \times 10^{-8}$ \\
\texttt{set-d-v-10.0-ns-0100} & 100 & 10.0 & 5826 & $0.42 \pm 0.01$ & $0.087 \pm 0.004$ & $0.0$ & $0.38 \pm 0.01$ & $0.0008 \pm 0.0004$ & $\left(1.19 \pm 0.04\right) \times 10^{-9}$ & $\left(3.81 \pm 0.15\right) \times 10^{-10}$ \\
\texttt{set-d-v-30.0-ns-0100} & 100 & 30.0 & 5824 & $0.48 \pm 0.01$ & $0.0017 \pm 0.0005$ & $0.0$ & $0.23 \pm 0.01$ & $0.0$ & $\left(1.86 \pm 0.07\right) \times 10^{-10}$ & $\left(5.52 \pm 0.24\right) \times 10^{-11}$ \\
\texttt{set-d-v-1.00-ns-0300} & 300 & 1.0 & 5825 & $0.0017 \pm 0.0009$ & $0.0012 \pm 0.0006$ & $0.0007 \pm 0.0003$ & $0.0017 \pm 0.0009$ & $0.0014 \pm 0.0007$ & $\left(6 \pm 4\right) \times 10^{-9}$ & $\left(5.1 \pm 3.0\right) \times 10^{-9}$ \\
\texttt{set-d-v-3.00-ns-0300} & 300 & 3.0 & 5812 & $0.35 \pm 0.03$ & $0.27 \pm 0.03$ & $0.16 \pm 0.02$ & $0.35 \pm 0.03$ & $0.28 \pm 0.03$ & $\left(3.9 \pm 1.0\right) \times 10^{-7}$ & $\left(1.8 \pm 0.6\right) \times 10^{-7}$ \\
\texttt{set-d-v-10.0-ns-0300} & 300 & 10.0 & 5807 & $0.48 \pm 0.02$ & $0.117 \pm 0.007$ & $0.0017 \pm 0.0006$ & $0.44 \pm 0.02$ & $0.0033 \pm 0.0008$ & $\left(4.68 \pm 0.30\right) \times 10^{-9}$ & $\left(1.64 \pm 0.12\right) \times 10^{-9}$ \\
\texttt{set-d-v-30.0-ns-0300} & 300 & 30.0 & 5802 & $0.55 \pm 0.02$ & $0.003 \pm 0.001$ & $0.0$ & $0.29 \pm 0.02$ & $0.0$ & $\left(7.7 \pm 0.5\right) \times 10^{-10}$ & $\left(2.39 \pm 0.18\right) \times 10^{-10}$ \\
\texttt{set-d-v-1.00-ns-0500} & 500 & 1.0 & 5824 & $0.002 \pm 0.002$ & $0.001 \pm 0.001$ & $0.0007 \pm 0.0007$ & $0.002 \pm 0.002$ & $0.002 \pm 0.002$ & $\left(1.3 \pm 1.2\right) \times 10^{-9}$ & $\left(4 \pm 4\right) \times 10^{-9}$ \\
\texttt{set-d-v-3.00-ns-0500} & 500 & 3.0 & 5824 & $0.35 \pm 0.04$ & $0.26 \pm 0.03$ & $0.15 \pm 0.02$ & $0.35 \pm 0.04$ & $0.27 \pm 0.04$ & $\left(3.5 \pm 0.6\right) \times 10^{-7}$ & $\left(1.4 \pm 0.4\right) \times 10^{-7}$ \\
\texttt{set-d-v-10.0-ns-0500} & 500 & 10.0 & 5815 & $0.48 \pm 0.03$ & $0.121 \pm 0.009$ & $0.0002 \pm 0.0002$ & $0.45 \pm 0.03$ & $0.004 \pm 0.002$ & $\left(8.1 \pm 0.6\right) \times 10^{-9}$ & $\left(3.0 \pm 0.4\right) \times 10^{-9}$ \\
\texttt{set-d-v-30.0-ns-0500} & 500 & 30.0 & 5814 & $0.55 \pm 0.02$ & $0.0041 \pm 0.0009$ & $0.0$ & $0.30 \pm 0.02$ & $0.0$ & $\left(1.30 \pm 0.08\right) \times 10^{-9}$ & $\left(4.14 \pm 0.33\right) \times 10^{-10}$ \\
\texttt{set-d-v-1.00-ns-1000} & 1000 & 1.0 & 5826 & $0.03 \pm 0.03$ & $0.03 \pm 0.02$ & $0.02 \pm 0.02$ & $0.03 \pm 0.03$ & $0.03 \pm 0.03$ & $\left(4 \pm 4\right) \times 10^{-7}$ & $\left(5 \pm 4\right) \times 10^{-7}$ \\
\texttt{set-d-v-3.00-ns-1000} & 1000 & 3.0 & 5836 & $0.51 \pm 0.03$ & $0.40 \pm 0.03$ & $0.23 \pm 0.03$ & $0.51 \pm 0.03$ & $0.42 \pm 0.04$ & $\left(5.9 \pm 0.7\right) \times 10^{-7}$ & $\left(2.5 \pm 0.5\right) \times 10^{-7}$ \\
\texttt{set-d-v-10.0-ns-1000} & 1000 & 10.0 & 5830 & $0.56 \pm 0.03$ & $0.16 \pm 0.02$ & $0.003 \pm 0.001$ & $0.54 \pm 0.03$ & $0.012 \pm 0.005$ & $\left(2.34 \pm 0.29\right) \times 10^{-8}$ & $\left(9.0 \pm 1.3\right) \times 10^{-9}$ \\
\texttt{set-d-v-30.0-ns-1000} & 1000 & 30.0 & 5814 & $0.64 \pm 0.02$ & $0.007 \pm 0.002$ & $0.0$ & $0.41 \pm 0.04$ & $0.0$ & $\left(3.8 \pm 0.4\right) \times 10^{-9}$ & $\left(1.34 \pm 0.20\right) \times 10^{-9}$

  \end{tabular}%
  }
\end{table*}

For the final set of simulations in this paper the intent was to determine how star-forming regions with higher densities with larger populations would affect disk enrichment.
In this set, simulations with a total star-forming region population of 100, 300, 500 or 1000 stars were introduced to an interloper of a velocity of either \SI{1}{km.s^{-1}}, \SI{3}{km.s^{-1}}, \SI{10}{\kms} or \SI{30}{km.s^{-1}}.
The interloper placed at a distance of \SI{3}{pc} from the cluster initially.
Repetition was performed for each parameter space combination, but the number of repetitions were scaled with the population size in order to ensure approximately the same number of total disks for each parameter space combination (100-star simulations were repeated 60 times, 300-star were repeated 20 times, 500-star repeated 12 times, and 1000-star simulations repeated 6 times).
There is some fluctuation in the total number of disks between sub-sets, this is due to varying numbers of high-mass stars between simulations.

Table \ref{tab:setd} details the averaged results of the set D simulations.
Enrichment quantity is found to vary due to velocity, as observed in previous simulation sets, what is notable is that there is not a significant difference in enrichment due to population size.
This is further expounded upon by Fig. \ref{fig:setd-isotopes}, where we can see that the only major difference between each population bin is the population of disks with Solar-like parent stars, due to the increased cluster population.
In the 1000-star sub-set, we see that $v_i = \SI{1}{\kms}$ enriched disks are present, despite the greater distance needed to travel from the start of the simulation.
This can be attributed to a larger number of stars at the periphery of the cluster, as well as the increased gravitational attraction of the more massive cluster.
Fig. \ref{fig:setd-com} shows the distance from the interloper to the nearest cluster star as a function of time, binned by velocity, we see that in the more massive cluster cases the interloper ``trapping'' seen in Fig. \ref{fig:set-bdistance-from-centre} also occurs. However, this would not affect the results of this set significantly as the limiting factor is the time required for the interloper to intersect the cluster. 
Overall, this further shows that interloper timing and velocity are the most important factors in causing efficient SLR enrichment, and an increased star-forming region population size does not significantly affect enrichment amount.

\section{Discussion}
\label{sec:discussion}

Based on our results we have determined that an interloping AGB star is an effective method of SLR enrichment, especially for the SLR \fe{}.
Whilst Wolf-Rayet star winds can produce the amount of \al{} enrichment present in the early Solar System, the disk disruption from these massive stars \citep{nicholsonRapidDestructionProtoplanetary2019,DaffernPowell22b,patelPhotoevaporationEnrichmentCradle2023a} and the long timescales to attain supernovae enrichment in \fe{} \citep{limongiPresupernovaEvolutionExplosive2018,2020ApJ...890...51E} would likely preclude the formation of giant planets (though WR winds and/or supernovae could still be a viable SLR source for planetary systems devoid of gas giants).

We have shown that the delivery of SLRs from AGB stars is capable of explaining the observed abundances in the early Solar system, whilst also preserving the protoplanetary disk as the AGB stars have much slower winds and limited FUV flux.
Our simulations produce higher-than-Solar \al{} concentrations, and frequently produces \fe{} enrichment above the higher-estimate value.
Whilst interlopers can cause significant amounts of enrichment, the amount of enrichment is dependent on a number of parameters.
Principally, interloper velocity and ensuring the interloping AGB star is experiencing a high rate of mass loss as it passes through the star-forming region are the most crucial variables observed.
In our earlier simulations enrichment was highly stratified by velocity, this was due to not adjusting the interlopers initial distance, $x_i$, from the cluster.
From these simulations we observed that above $v_i = \SI{10}{\kms}$ enrichment to Solar-System levels becomes unlikely, though this is still within the expected range of relative velocities between nearby objects in the galaxy.
However, our later simulations which varied $x_i$ and $y_i$ significantly show that with a large enough initial distance to the star-forming region, large numbers of enriched disks can be produced again, even with interloper velocities approaching \SI{30}{km.s^{-1}}.
We can infer from this that the initial time before intersection is more important, and that in the case of higher-velocity interlopers, the amount of disks that undergo significant enrichment is actually quite insensitive to the initial interloper position -- provided that it has reached a later period of the AGB phase where mass-loss rates are higher.
With a suitable initial distance or time before interaction, even ``runaway'' interlopers  ($v_i > \SI{30}{km.s^{-1}}$) can significantly enrich disks in star-forming regions.

The initial mass or population of the cluster does not significantly affect results, aside from gravitationally, where AGBs are accelerated towards the star-forming region at a greater rate if the mass of the star-forming region is higher.
In some of our higher-mass star-forming regions the interloping AGB star can be trapped, and while this does cause increased amounts of enrichment, the higher-mass regions are also more likely to contain massive stars \citep{Parker07,nicholsonSupernovaEnrichmentPlanetary2017} which would destroy the protoplanetary disks \citep{concha-ramirezExternalPhotoevaporationCircumstellar2019,nicholsonRapidDestructionProtoplanetary2019}. In order to trap AGB stars in low-mass ($N_\star = 100$) star-forming regions, interloper velocities $<\SI{1}{\kms}$ are required.

Whilst AGB interlopers could be a potential mechanism of SLR enrichment, a few outlying questions remain.
In particular, the probability of these interloping events even occurring has not been ascertained, a single example exists in literature, while there are other potential interloper candidates, no wide-scale survey for these interlopers has been performed.
We also observe that ``near-misses'' from interlopers, which come within $2\,r_c$ of the star-forming region can still cause enrichment, as such, as long as the interloper passes through or near the star-forming region, enrichment can occur.
This increases the probability of successful enrichment, as the star does not have to pass directly through the star-forming region.
Future data releases from Gaia could provide additional examples of interlopers, and may also reveal the presence of trapped AGB interlopers.
An outlying concern with our current model are the presence of some free parameters in our simulations.
Using a large -- though still probable -- disk size would result in larger enrichments amounts; disks on the order of \SI{100}{AU} would result in an approximately order-of-magnitude change, still enough for a few highly-enriched disks.
Uncertainties also exist in the efficiency of core-produced SLRs such as \fe{} being transferred to the winds, though this would reduce \fe{} enrichment closer to the \citet{tangAbundanceDistributionOrigin2012} low-estimate value of $Z\rms{60Fe} = 10^{-8}$.
Additionally, modelling dissipation of the AGB wind over time would result in a lower amount of enrichment, while SLR decay within the ``tunnel'' of the AGBs previous path is accounted for, expansion of the region as the wind outflows would decrease the density of the region, resulting in a lower rate of absorption of material.
The latter issue could be addressed in a subsequent model that utilises an SPH-based outflow model from the AGB star, and would not require the use of interaction regions.

\section{Conclusions}
\label{sec:conclusion}

In this paper we performed a large parameter space search using an $N$-body model to explore how the combined initial conditions of a star-forming region and interception properties of an interloping AGB star affect the enrichment of protoplanetary disks with short-lived radioisotopes. Our conclusions are the following:

\begin{enumerate}
  \item We found that interloping AGB stars offer an abundant source of both \al{} and \fe{}, and can enrich multiple disks in a low-mass star-forming region to Solar System amounts in a short period of time (several Myr), absent of locally produced massive stars and supernovae.
  \item Interloper enrichment is sensitive to certain parameters, in particular the time at which the AGB begins interacting with the star-forming region, as well as its initial velocity.
  \item It appears that in the case of low-velocity ``walkaway'' interlopers, the interloper can be captured by the star-forming region, rather than simply passing through it. Whilst this event produces marginally more enrichment from the AGB, it is less likely, and is not a requirement for high (Solar system-like) levels of enrichment.
\end{enumerate}

Future work should focus on determining the overall probability of star-forming regions undergoing an interloper encounter.
In addition, observational searches for further interloping (and/or trapped AGB stars) in star-forming regions would help determine the likelihood of Solar system (and other planetary system) enrichment from AGB stars.
AGB stars, being the primary source of interstellar dust in the galaxy, have an outsized impact on the interstellar medium, and their being responsible for another facet of planetary evolution is well worth further investigation.

\section*{Acknowledgements}

JWE and RJP acknowledge funding from the Royal Society, in the form of a Dorothy Hodgkin research fellowship award to RJP.

\section*{Data availability}

The data underlying this article will be shared on reasonable request to the corresponding author.

\bibliographystyle{mnras}
\bibliography{references}

\bsp	%
\label{lastpage}
\end{document}